\documentclass[lettersize,journal]{IEEEtran}
\usepackage{amsmath,amsfonts}
\usepackage{algorithmic}
\usepackage{algorithm}
\usepackage{array}
\usepackage{acronym}
\usepackage[caption=false,font=normalsize,labelfont=sf,textfont=sf]{subfig}
\usepackage{xcolor}
\usepackage{soul}
\usepackage{textcomp}
\usepackage{stfloats}
\usepackage{url}
\usepackage{verbatim}
\usepackage{graphicx}
\usepackage{multirow}
\usepackage{cite}
\acrodef{3GPP}{3rd Generation Partnership Project}
\acrodef{AI}{Artificial Intelligence}
\acrodef{AIoT}{Ambient-IoT}
\acrodef{AP}{Access Point}
\acrodef{CN}{Core Network}
\acrodef{CNN}{Convolutional Neural Network}
\acrodef{CPU}{Central Processing Unit}
\acrodef{FL}{Federated Learning}
\acrodef{FLOP}{Floating Point Operations}
\acrodef{FLOPS}{Floating Point Operations Per Second}
\acrodef{GFLOPS}{Giga Floating Point Operations Per Second}
\acrodef{GO}{Goal-Oriented}
\acrodef{GOPS}{Giga Operations Per Second}
\acrodef{GPU}{Graphics Processing Unit}
\acrodef{ILP}{Integer Linear Programming}
\acrodef{IIoT-C}{Intelligent IoT-Cloud aided}
\acrodef{IIoT-D}{Intelligent IoT-Device only}
\acrodef{IoT}{Internet of Things}
\acrodef{ITU}{International Telecommunication Union}
\acrodef{KPI}{Key Performance Indicator}
\acrodef{MO}{Multi-Objective}
\acrodef{NB}{NarrowBand}
\acrodef{NF}{Network Function}
\acrodef{NR}{New Radio}
\acrodef{RAN}{Radio Access Network}
\acrodef{RF}{Radio Frequency}
\acrodef{RL}{Reinforcement Learning}
\acrodef{SNR}{Signal-to-Noise Ratio}
\acrodef{TIoT}{Traditional IoT}
\acrodef{UE}{User Equipment}
\acrodef{UL}{Uplink}
\acrodef{UPF}{User Plane Function}
\acrodef{YOLO}{You Only Look Once}

\begin{document}

\title{A Goal-Oriented Networking Approach\\for Intelligent IoT Service Deployment}


    
    

\author{Federico Tonini, Davide Borsatti, Wint Yi Poe, Riccardo Trivisonno, Walter Cerroni
\thanks{F. Tonini is with CNIT WiLab, Bologna, Italy (e-mail: federico.tonini@wilab.cnit.it). \textit{Corresponding author.}}
\thanks{D. Borsatti, W. Cerroni are with DEI, University of Bologna, Italy and CNIT WiLab, Bologna, Italy (e-mail: davide.borsatti@unibo.it; walter.cerroni@unibo.it).}
\thanks{W. Y. Poe, R. Trivisonno are with Huawei Heisenberg Research Center, Munich, Germany (e-mail: wint.yi.poe@huawei.com; riccardo.trivisonno@huawei.com).}
}



\maketitle

\begin{abstract}
The first 6G standardization efforts are about to start, shaping the new generation of mobile networks. The IMT-2030 extends the IMT-2020 by expanding its usage scenarios to Immersive, Massive, and Hyper-Reliable and Low-Latency Communications. It also introduces novel scenarios by integrating Artificial Intelligence and Sensing with Communication and supporting Ubiquitous Connectivity. Compared to the previous generation, 6G is expected to improve not only throughput and latency, but also coverage and energy efficiency.
A paradigm called Goal-Oriented (GO) communications has recently emerged as a promising solution to improve network efficiency. It relies on the fact that the goal of the communication network is to achieve a specific task with a defined accuracy, rather than creating perfect data delivery. Intelligent devices can pre-process data to send only what is relevant to achieve the task, thus saving precious network resources and energy. Recent works demonstrate that incorporating service- and application-level KPIs in the network allows to achieve higher communication efficiency for devices, but the consequence of using such techniques on the network itself has not yet been explored.
This paper proposes a practical end-to-end framework to assess energy consumption, latency, and goal accuracy KPIs, which includes a Multi-Objective optimization model to evaluate the trade-offs between the multiple KPIs relevant to GO networking.
We demonstrate, through simulation, that the network can benefit from the application of the GO paradigm, indicating its potential in future network architectures.
\end{abstract}


\section{Introduction}

\IEEEPARstart{A}{fter} half a decade from the first 5G network deployments, the development of 6G is undergoing. Even though there is no official definition of 6G yet, it is general consensus that 
6G will provide services beyond traditional communication \cite{ngmn-6g,hexa-x2-final}. \ac{AI} will play a major role in the future generation of networks, where new services will be enabled by distributed and ubiquitous \ac{AI} capabilities, such as the one provided by the deployment of intelligent \ac{IoT} devices and edge-cloud continuum \cite{survey-IIoT}. This, in turn, will lead to a convergence of communication and computation, where \ac{AI} and communication networks will be tightly integrated. 6G is also expected to be sustainable by design, reaching unprecedented levels of energy efficiency and reduced resource consumption, thus contributing to meet drastic global emissions reductions. 
Furthermore, 6G is expected to evolve into a more user‑centric network \cite{ITU-M2160}, enabling user‑focused communications and dynamic service delivery tailored to individual needs \cite{CalCN21}. 
New paradigms that break the traditional separation between communication and computation are required to achieve these goals.

\ac{GO} communications represent a relatively new approach that prioritizes the transmission of only the most relevant data necessary to achieve a specific task or goal, which inherently leads to more efficient use of resources, including energy \cite{CalCN21}. Instead of focusing on delivering all bits with a required quality, \ac{GO} communications leverage intelligence at the \ac{IoT} devices and service level information to reduce the amount of data to be sent through the network while achieving the pre-defined goal with a specified accuracy level. Recent solutions for intelligent \ac{IoT} adopting \ac{GO} communications mostly focus on reducing energy consumption during transmission or optimizing communication and computing resources, targeting \acp{KPI} for single \ac{IoT} devices or cloud computing nodes.

However, to the best of our knowledge, what is still missing is an evaluation methodology that allows to take into account also the effects on other network components, considering the impact of \ac{GO} communications from an \textit{end-to-end} perspective. Such a \textit{\ac{GO} networking} perspective is, in our opinion, what must be considered to assess user-centered, service-level \acp{KPI} to determine how satisfactorily an intended goal has been achieved. Interestingly, the introduction of service-level \acp{KPI} along with conventional network \acp{KPI} might generate conflicting situations. For example, this is the case of a system that aims to reduce energy consumption by running low-complexity (and low-performing) \ac{AI} algorithms, while also seeking to maximize the overall acceptable accuracy of the intended goal. In such cases, \ac{MO} optimization problems with conflicting objectives must be solved to allocate the right network resources and distribute the most appropriate \ac{AI} models.

In this paper, we propose a novel optimization framework for deploying \ac{AI} workloads in \ac{IoT} services with intelligence at the device level, following a \ac{GO} networking approach. We consider an \ac{IoT} scenario with intelligent cameras for object detection, where the \ac{AI}-based processing can be performed at the device, in the cloud, or in a hybrid configuration. In particular, the main contributions of our work include:
\begin{itemize}
    \item The original formulation of a \ac{MO} optimization model based on \ac{ILP}, encompassing both service and network requirements, aiming at minimizing the energy consumption while achieving pre-defined service goals.
    \item The presentation of an intelligent \ac{IoT} use case on distributed \ac{AI}-based object detection, leveraging device sensing through cameras and different \ac{GO} networking strategies, with related \acp{KPI}.
    \item The performance evaluation of the proposed strategy under the previously mentioned use case with an end-to-end \acp{KPI} model for energy consumption, latency, and goal accuracy.
\end{itemize}

Compared to conventional \ac{IoT}, where all data are always sent to the cloud, we show that a \ac{GO} networking strategy with a proper selection of the right \ac{AI} model, parameters, and location in the network for the inference execution can lead to up to $60\%$ or $80\%$ savings depending on the imposition of strict or not-so-strict accuracy requirements, respectively.

The remainder of the paper is as follows. Section \ref{sec:rel-work} provides an overview of existing networking strategies for \ac{GO} communication and intelligent \ac{IoT} as well as standardization initiatives on the topic. Section \ref{sec:use-case} describes the use case and relevant \acp{KPI} required to assess the performance of the different strategies for intelligent \ac{IoT}. Section \ref{sec:optimization} describes the novel optimization model for Intelligent \ac{IoT} service deployment, while Section \ref{sec:results} reports a detailed analysis of the achieved performance and relevant trade-offs. Finally, Section \ref{sec:conclusion} concludes the paper.

\section{Efforts on IoT and GO Communications Towards 6G}
\label{sec:rel-work}

This section first provides an overview of the state of the art in \ac{GO} communications for Intelligent \ac{IoT}. Then, an overview of the main standardization efforts and corresponding relevant technical areas considered in the 6G standardization process is presented.

\subsection{State-of-the-art of GO Communications Strategies and Intelligent IoT}

\ac{GO} communication can be defined as a communication paradigm where the primary objective is the fulfillment of a specific task or goal, characterized by a set of requirements that, if attained, determine its accomplishment. This approach emphasizes the relevance and significance of the transmitted information in relation to the intended outcome, rather than merely focusing on the accurate transmission of data. In this context, the effectiveness of the communication is measured by how well it supports the achievement of the predefined goal, which can include various performance metrics such as accuracy, latency, and resource efficiency \cite{DilIOT23}. While traditional communication systems focus on making sure that messages are sent and received accurately, without considering the purpose of the communication, in \ac{GO} communications the communication system should be designed to help achieve specific tasks, rather than just ensuring that the messages are reconstructed perfectly \cite{GO-2}. This behavior has direct implications for the air interface by reducing traffic load, improving spectrum usage, and enabling more efficient access strategies, especially important in scenarios involving low-end, energy-constrained devices.

\ac{GO} communications are gaining increasing attention in the research community, with a growing number of recent works emerging on the topic.
%
The authors in \cite{BinTGCN23} employ a convolutional encoder that is jointly trained with a convolutional decoder at the edge server to minimize the amount of data sent over the network, which directly contributes to lower energy consumption during transmission. A similar approach is reported in \cite{GutPIMRC23}, showing the benefits of reducing the amount of data that needs to be transmitted while simultaneously improving robustness against communication channel errors and compensating for poor \ac{SNR}.
In \cite{DilIOT23}, the authors propose to equip \ac{IoT} devices with a \ac{GO} communication module that distills and transmits only the relevant data features necessary to achieve the communication goal. This module employs effective sensing techniques to ensure that only pertinent information is collected, thereby optimizing resource usage. The framework also includes a \ac{GO} optimization engine that dynamically manages network resources based on real-time measurements of energy consumption and goal effectiveness. The paper provides different examples of practical applications of their approach in scenarios such as edge inference, cooperative sensing, and \ac{FL}, highlighting the substantial gains in energy efficiency, latency, and accuracy achieved through \ac{GO} communication strategies.

The solution provided in \cite{MerEUCNC22} employs adaptive resource management strategies that dynamically allocate communication and computational resources based on the current system state and application requirements. The use of ensemble inference, where multiple models work collaboratively to improve accuracy and efficiency, enhances energy efficiency compared to standalone inference methods, as it allows relaxation of the communication requirements for the \acp{UE}. Other techniques can be used to reduce the required transmitted bits while effectively achieving communication goals. In \cite{GO-2}, the authors analyze \ac{GO} signal preprocessing and quantization techniques that can operate on the input signals and data by focusing on relevant features and removing unnecessary data, thus reducing the amount of information that needs to be transmitted. Data clusters can also be formed, e.g., in a scenario where \ac{IoT} devices are tasked with monitoring environmental conditions, so that they can transmit only significant changes rather than continuous data streams.

All the aforementioned works contribute significantly to advancing energy-efficient solutions for \ac{GO} communications. However, they primarily focus on the energy consumption and latency of \ac{IoT} devices and cloud nodes, failing to consider the contribution of the rest of the network. Studies indicate that the \acf{RAN} alone is responsible for $70 \%$ of a mobile network energy consumption \cite{gsma-ran}, and its contribution cannot be neglected when evaluating \ac{GO} networking strategies. A true end-to-end analysis of \acp{KPI} such as energy consumption, latency, and task accuracy must be provided to support the adoption of existing and the development of new \ac{GO} strategies. 
Our prior work \cite{ENS25} introduced one of the first comprehensive end-to-end \ac{KPI} models capturing energy consumption, latency, and accuracy for \ac{IoT} applications. Building on this, we advance the state of the art by proposing a novel \ac{MO}-\ac{ILP}-based optimization framework for joint \ac{AI} model selection and network resource allocation to achieve the service goal, and introduce an epsilon-constrained approach to systematically explore the trade-offs among these \acp{KPI}.

\subsection{Standardization Initiatives and 
Relevance}

\ac{ITU} provided a Recommendation containing usage scenarios and capabilities of IMT-2030 \cite{ITU-M2160}. While scenarios such as immersive communication, massive communication, and hyper-reliable and low-latency communication represent a direct evolution of IMT-2020 services, \ac{ITU} also introduces integrated sensing and communication, \ac{AI} and communication, and ubiquitous connectivity as new types of services envisioned for 6G. Among others, the Recommendation highlights the need to support distributed computing and \ac{AI} applications as well as the integration of \ac{AI} and compute functionalities into IMT-2030. Sensing, \ac{AI}, and sustainability are notably among the new capabilities that IMT-2030 is expected to provide.

The \ac{3GPP} has also recently started its efforts in the definition of technical specifications for 6G.
Among the potential areas that will lay the foundations of future generation networks, the support for different \ac{IoT} device types (e.g., for \ac{AIoT} \cite{tr38769}), and service-aware intelligent networks have been considered, along with the native integration of \ac{AI} frameworks \cite{3gpp-ws-6G} for intelligent \ac{IoT} services.
6G capabilities will also allow embodiment of sensing with \ac{RF} or other non-\ac{RF} sources (e.g., environmental detectors, cameras, etc.) \cite{etsiISC001}. \ac{IoT} devices can therefore contribute to enriching the network sensing capabilities.

The support for \ac{IoT} communications started with Release 16 with the introduction of cellular \ac{IoT} and its evolution for the 5G System, providing also enhancements for \ac{NB}-\ac{IoT} \cite{tr21916}. Release 17 extends the support to non-terrestrial networks, industrial \ac{IoT} while introducing energy efficiency and power saving mechanisms \cite{tr21917}. Release 18 introduces support for personal \ac{IoT} and residential networks \cite{tr21918}.
Release 19 introduces \ac{AIoT} for \ac{IoT} applications with ultra-low cost and ultra-low power devices \cite{tr38769}.
The 6G System is expected to support diverse \ac{IoT} device types and use cases by design from day one \cite{3gpp-ws-6G}.
This calls for mechanisms that can efficiently manage intelligent, \ac{AI}‑enhanced \ac{IoT} devices and coordinate their capabilities within the network \cite{survey-IIoT}.

In this paper, we focus on a use case where \ac{IoT} devices are able to sense the environment by taking images, which are elaborated by \ac{AI} models integrated and distributed in both the devices and the network. We propose a strategy to deploy such services on the network in an energy-efficient way, while providing the necessary service guarantees to achieve the communication goal. We prove by simulation the benefits provided by the integration of intelligent \ac{IoT} devices and services in the network in terms of energy efficiency and resource usage.



\section{Use Case and KPIs description}
\label{sec:use-case}

\begin{figure}[t]
    \centering
    \includegraphics[width=\linewidth]{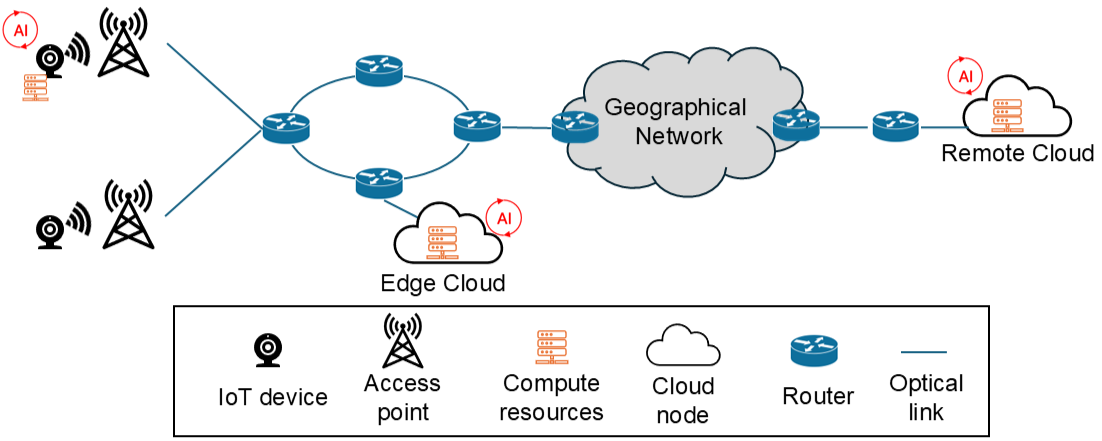}
    \caption{Example of an end-to-end network with intelligent IoT devices.}
    \label{fig:sample-scenario}
\end{figure}

In \ac{IoT} contexts, a common use case for \ac{GO} communications is object detection \cite{BinTGCN23,GutPIMRC23,DilIOT23,MerEUCNC22}, where the goal of the communication system is to detect objects in images.
A possible scenario for intelligent \ac{IoT} is depicted in Figure \ref{fig:sample-scenario}. An\ac{IoT} device must be deployed to capture environmental data in its area, like images of the surroundings. To do so, one must decide on the specific \ac{AI} model and where to execute inference tasks based on energy, latency, and accuracy requirements. Two types of devices are considered. Conventional \ac{IoT} devices rely on the network resources to deliver data and on \ac{AI} models on the cloud to perform inference tasks. Instead, Intelligent \ac{IoT} devices can perform data processing locally with \ac{AI} models, with or without network assistance, depending on targets like latency, energy or task efficiency, and accuracy. A set of \acp{AP}, optical links and devices, and routers interconnects devices and edge/remote cloud nodes.
In the following, we first present the \ac{IoT} strategies for \ac{GO} Networking. Then, the \acp{KPI} modelling is proposed and discussed.

\subsection{Intelligent IoT Strategies for GO Networking}

\begin{figure*}[!t]
\centering
\subfloat[TIoT.]{\fbox{\includegraphics[width=0.3\textwidth]{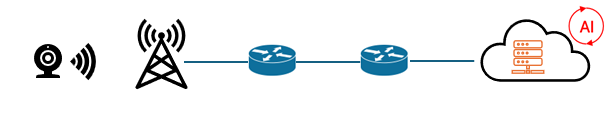}}%
\label{fig:TIoT}}
\hfil
\subfloat[IIoT-D.]{\fbox{\includegraphics[width=0.3\textwidth]{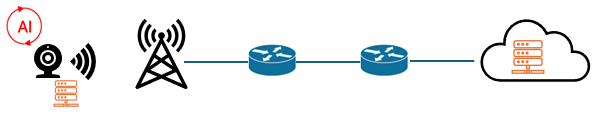}}%
\label{fig:IIoT-D}}
\hfil
\subfloat[IIoT-C.]{\fbox{\includegraphics[width=0.3\textwidth]{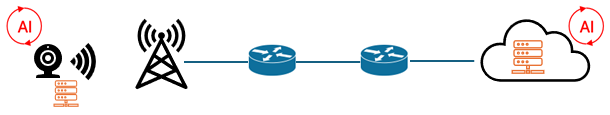}}%
\label{fig:IIoT-C}}
\caption{Examples of considered scenarios, showing where the intelligence is used.}
\label{fig:cases}
\end{figure*}

We assume \ac{IoT} devices to be equipped with cameras and deployed in the target areas to capture images, which are then analyzed by \ac{AI} models based on neural networks to detect objects.
We define the following strategies:
\begin{itemize}
    \item \textit{\ac{TIoT}}: \ac{IoT} devices equipped with cameras periodically send captured images to cloud nodes, where object detection and recognition tasks are performed. Figure \ref{fig:TIoT} depicts an example of this approach. This represents a traditional \ac{IoT} scenario, where each device continuously sends images to the cloud and the inference is executed for all the data. In this case, network energy and resources are consumed to send data, and a certain latency is introduced as data is not immediately processed at the source. The images are processed in the cloud, where efficient hardware is typically available, making the inference task both energy- and time-efficient.
    

     \item \textit{\ac{IIoT-D}}: \ac{IoT} devices equipped with cameras leverage the on-board intelligence to execute inference tasks. In this case, the network does not introduce energy consumption or latency, as the data is processed at the source. However, most \ac{IoT} devices are equipped with limited computational capabilities or constrained by their battery life, and their accuracy, energy consumption, and latency are often affected. For example, performing a complex inference task with a large number of operations may be latency- and energy-inefficient on low-end and inexpensive \ac{IoT} devices. An example of this approach is reported in Figure \ref{fig:IIoT-D}.
    
    \item \textit{\ac{IIoT-C}}: intelligent \ac{IoT} devices, equipped with cameras and compute capabilities, are able to perform simple inference tasks using limited-size models in order to filter out irrelevant images. Images are only sent to the cloud for analysis when a specific object is detected with a specified likelihood (referred to as \textit{threshold}). More powerful \ac{AI} models are usually available in the cloud for more accurate object detection. With this approach, only the meaningful data to achieve the goal is transported over the network, thus saving energy compared to TIoT. Compared to IIoT-D, IIoT-C reduces latency and energy consumption, as only simple models are executed at the \ac{IoT} level, while more complex models are executed on efficient hardware platforms in the cloud. An example is reported in Figure \ref{fig:IIoT-C}.

\end{itemize}

While all three strategies achieve the goal of detecting the presence of an object in the image, they may perform differently in terms of end-to-end energy consumption, latency, and accuracy of the task. Therefore, we need a formal definition of these three \acp{KPI}.

\subsection{End-to-End KPIs for GO networking}
\label{sec:kpis}

To assess the performance of the different \ac{GO} strategies, we leverage the \ac{KPI} modelling that we originally introduced in \cite{ENS25}, for which we only provide a summary and defer the readers to \cite{ENS25} for further details. In the model, we use the terms \ac{UE} and \ac{IoT} device interchangeably. We specifically focus on the User Plane energy consumption in \ac{UL}, with \ac{IoT} devices that might perform object detection or send data to a computation node where the inference is executed. We consider conventional networking \acp{KPI}, such as energy consumption and latency, but also include in the modelling an application-oriented \ac{KPI} like the accuracy, a typical \ac{GO} metric.

\subsubsection{Energy consumption}

the energy consumption is expressed per \ac{UE} and is modeled as follows:
\begin{equation}
\begin{aligned}
E = & \, E_{\text{UE}}^{\text{proc}} + E_{\text{UE}}^{\text{trans}} + E_{\text{UE,AP}}^{\text{recv}} + E_{\text{UE,AP}}^{\text{proc}} + \\
& + E_{\text{UE,TN}} + E_{\text{UE,UPF}} + E_{\text{UE,Comp}}^{\text{proc}}.
\end{aligned} \label{eq:energy}
\end{equation}

The \ac{UE} processing ($E_{\text{UE}}^{\text{proc}}$) and the compute node processing ($E_{\text{UE,Comp}}^{\text{proc}}$) terms
account for the energy consumed by the inference tasks performed onboard each intelligent \ac{UE} and in the cloud nodes, respectively. They are assumed to be \cite{BinTGCN23} \cite{MerEUCNC22}:
\begin{equation}
E_{\text{UE}}^{\text{proc}} = E_{\text{UE,Comp}}^{\text{proc}} = \kappa f^3 \tau
\label{eq:UE_proc_consumption_v2}
\end{equation}
where $\kappa$ is the effective switched capacitance of the processor, $f$ is the \ac{CPU} maximum clock frequency, and $\tau$ is the inference time or, equivalently, the ratio between the number of \ac{FLOP} for the inference task and the number of \ac{FLOPS} of the \ac{CPU} \cite{BinISWCS24}.

The \ac{UE} data transmission term ($E_{\text{UE}}^{\text{trans}}$)
accounts for the energy spent by the \ac{UE} to send data to the \ac{AP} and can be expressed as:

\begin{equation}
E_{\text{UE}}^{\text{trans}} = \frac{d}{\alpha} P_t
\end{equation}

where $d$ is the amount of data transmitted at an average rate $\alpha$, and $P_t$ is the transmitted power of the \ac{UE}.

The energy spent at the \ac{AP} in reception of \ac{UE} data ($E_{\text{UE,AP}}^{\text{recv}}$)
is assumed to be proportional to the \ac{UE} data volume transmitted during \ac{AP} reception time $t_r$, multiplied by the power consumed by the \ac{AP} to receive signals ($P_{\text{AP}}^{\text{recv}}$) and by the power amplifiers ($P_{\text{PA}}$), plus an overhead factor $\eta$ accounting for AC/DC conversion and cooling:

\begin{equation}
\label{eq:E_AP_recv}
E_{\text{UE,AP}}^{\text{recv}} = \frac{\text{UE}_{\text{data volume}}}{\text{AP}_{\text{tot data volume}}} \cdot t_{r} \cdot (P_{\text{AP}}^{\text{recv}} + P_{\text{PA}}) \cdot (1+\eta).
\end{equation}

$P_{\text{AP}}^{\text{recv}}$ depends on the number of antennas and sectors, bandwidth, load, and spectral efficiency of the \ac{AP} \cite{power-model-future-proof}.

The energy required by the \ac{AP} digital processing for the considered \ac{UE} ($E_{\text{UE,AP}}^{\text{proc}}$)
consists of the energy consumption of all the \ac{AP} \acp{NF} \cite{ts28554}. Similarly to \eqref{eq:E_AP_recv}, it can be expressed as:

\begin{equation}
E_{\text{UE,AP}}^{\text{proc}} = \frac{\text{UE}_{\text{data volume}}}{\text{AP}_{\text{tot data volume}}} \cdot t_{r} \cdot P_{\text{AP}}^{\text{proc}} \cdot (1+\eta)
\end{equation}

where $P_{\text{AP}}^{\text{proc}}$ is the power consumed for the baseband processing at the \ac{AP} and depends on the \ac{GOPS} needed to perform each \ac{NF} \cite{power-model-future-proof}.

The per-\ac{UE} energy consumption of the transport network ($E_{\text{UE,TN}}$)
is assumed to be proportional to the amount of data generated by the \ac{UE} ($\text{UE}_{\text{data volume}}$) over the used links of the transport network, the rate of the optical interface of each link ($\text{Link}_\text{rate}$), and the power consumed by a couple of optical transponders ($P_{\text{tr}}$) and network interface cards ($P_{\text{nic}}$) :

\begin{equation}
E_{\text{UE,TN}} = \sum_{\text{used-links}} \frac{\text{UE}_{\text{data volume}}}{\text{Link}_\text{rate}} \cdot 2 \cdot (P_{\text{nic}} + P_{\text{tr}}).
\end{equation}

The \ac{UPF} processing energy consumption ($E_{\text{UE,UPF}}$) depends on the amount of \ac{UE} data in the observation period $\text{UE}_{\text{data volume}}$ over the data volume processed by the \ac{UPF} in the same period, and the energy consumed by the \ac{UPF}:

\begin{equation}
E_{\text{UE,UPF}} = \frac{\text{UE}_{\text{data volume}}}{\text{UPF}_{\text{tot data volume}}} \cdot \int\limits_{T_\text{observation}} P_{\text{UPF}}(t) \,dt
\end{equation}

where $P_{\text{UPF}}(t)$ represents the instantaneous \ac{UPF} power consumption, which is derived from real measurements in \cite{power-cons-meas}:

\begin{equation}
P_{\text{UPF}}(t) = 10.625 \cdot \text{UPF}_{\text{data rate}}(t) + 8500.
\end{equation}

If more \acp{UPF} are processing \ac{UE} data, the overall energy consumption is the summation of the contributions of each \ac{UPF}.

\subsubsection{Latency}
we define latency as the time elapsed from when the data becomes available at the \ac{UE} (or \ac{IoT} device) to when the application is ready to return a response after data processing. Latency is expressed as sum of different contributions:
\begin{equation}
L = L_{\text{radio}} + L_{\text{transport}} + L_{\text{routing}} + L_{\text{processing}}. \label{eq:lat}
\end{equation}
The \( L_{\text{radio}} \) represents the \ac{RAN} delay, i.e., the time needed to transmit data $d$ at rate $\alpha$ to the \ac{AP}:
%
\begin{equation}
L_{\text{radio}} = d / \alpha.
\end{equation}
The \( L_{\text{transport}} \) includes all the delays of the optical transport network. It is the sum of the delay introduced by the propagation over the fiber cables and the switching time ($t_{sw}$) introduced by each of the $N_{sw}$ switches along the path:
\begin{equation}
L_{\text{transport}} = \frac{\text{distance}}{2 * 10^{8}} + N_{sw}*t_{sw}.
\end{equation}
The \( L_{\text{routing}} \) is the delay introduced by $N_\text{UPF}$ \acp{UPF} along the path, each introducing a fixed delay $L_\text{UPF}$ for routing traffic:
\begin{equation}
L_{\text{routing}} = N_{\text{UPF}} * L_{\text{UPF}}.
\end{equation}
The \( L_{\text{processing}} \) is the data processing time for the application. It is given by the ratio between the number of operations required by the inference model divided by the \ac{GFLOPS} provided by the \ac{CPU}:
%
\begin{equation}
L_{\text{processing}} = \frac{\text{model operations}}{\text{GFLOPS CPU}}. \label{eq:lat-proc}
\end{equation}

\subsubsection{Accuracy}
\label{sec:acc}
From a system-level perspective, the goal is achieved if a task is performed with acceptable accuracy. Therefore, an application-level \ac{KPI} needs to be introduced to correctly assess the performance of the \ac{GO} strategy. The widely used $F_1$ Score is adopted as the accuracy metric:
\begin{equation}
F_1 = 2 \cdot \frac{\text{P} \cdot \text{R}}{\text{P} + \text{R}}.
\label{eq:f1}
\end{equation}
The precision ($P$) is the ratio of true positive detections to the total number of positive detections:

\begin{equation}
P = \frac{\text{True Positives (TP)}}{\text{True Positives (TP)} + \text{False Positives (FP)}}.  
\end{equation}

The recall ($R$) is the ratio of true positive detections to the total number of actual objects in the ground truth:

\begin{equation}
R = \frac{\text{True Positives (TP)}}{\text{True Positives (TP)} + \text{False Negatives (FN)}}. 
\end{equation}

When dealing with a cascade of models that progressively filter out unnecessary data (e.g., \ac{IIoT-C}), all models contribute to FN. Since relevant data are analyzed at the end of the cascade, FP and TP are accounted for only at the last model.
Finally, we present the complement of accuracy, the inaccuracy, which will be employed in the next Sections. By considering that the $F_1$ Score assumes values in the interval $[0, 1]$, the inaccuracy can be defined as:

\begin{equation}
\overline{F_1} = 1 - F_1.
\label{eq:f1-inacc}
\end{equation}

\section{GO Intelligent IoT Service Deployment}
\label{sec:optimization}

To deploy \ac{IoT} services for object detection, a \ac{GO} networking framework, aware of the service goals, is required to select the most appropriate set of resources on the network. This is the case of \ac{TIoT} and \ac{IIoT-C}.
Let us consider the sample network in Figure \ref{fig:sample-scenario}.
When \ac{IoT} devices need to send data to the cloud for processing, data travels through the \ac{AP} and reaches a cloud node via the transport network, routed through one or more \acp{UPF} along the path. Among all possible paths connecting an \ac{AP} to a cloud node, we assume that the shortest path is selected. Therefore, it is possible to pre-compute the shortest paths for each \ac{AP}-cloud node pair ($P$). On the cloud, inference tasks can be executed on any cloud node $c$ among the $C$ cloud locations. A model $m$ on the cloud must be selected among a set of available models ($M$), each characterized by a different amount of required data, computational complexity, and accuracy ($a_{m}$). While the accuracy in \ac{TIoT} case is provided only by the model deployed at the cloud, in the \ac{IIoT-C} case, the accuracy depends on the combination of models (see Figure \ref{fig:cases}).
The resources required to execute model $m$ on cloud node $c$, $r_{m,c}$, must not exceed the available resources $R_{c}$. Running inference model $m$ on compute node $c$ requires an energy cost $e_{m,c}$ and a processing time $l_{m,c}$, corresponding to $E_{\text{UE,Comp}}^{\text{proc}}$ and $L_{\text{processing}}$ in \eqref{eq:lat}, respectively.
The latency introduced by a path $p \in P$, denoted as $l_{m,p}$, depends on the chosen model $m$ since the data size to be sent varies with different inference models. This quantity corresponds to $L_{\text{radio}} + L_{\text{transport}} + L_{\text{routing}}$ in \eqref{eq:lat}. Similarly, the energy consumption of path $p$, denoted as $e_{m, p}$, includes all the energy consumption terms in \eqref{eq:energy} except those related to performing the inference task.

Given the service goals, the required and available network resources, we must find the most suitable set of resources to achieve the goal.
An optimization problem is formulated to determine (i) the inference model configuration, (ii) the compute node on the cloud for execution, and (iii) the path connecting the \ac{IoT} device to the compute node. This selection must account for the energy required to execute the model and the energy needed for data transfer, as well as latency over different paths, an estimation of which can be provided by data analytics functions of the Core Network. The accuracy values can be obtained, e.g., by testing each model on a labeled dataset and depend on data quality, model parameters and training.
In the following, we assume the minimization of the three conflicting \acp{KPI} introduced in Section \ref{sec:kpis}: \textit{energy consumption}, \textit{latency}, and \textit{inaccuracy}. We introduce two variables $x_{m,c}$ and $y_{m,p}$. $x_{m,c}$ is a binary variable equal to $1$ if configuration model $m \in M$ is selected to be executed at compute node $c \in C$; $0$ otherwise. $y_{m,p}$ is also a binary variable and it is equal to $1$ if path $p \in P$ is selected to carry traffic for the considered \ac{IoT} device using configuration model $m \in M$; $0$ otherwise. These two decision variables store the solution of the optimization problem. A summary of the notation is reported in Table \ref{tab:ILP-v2}.

\subsection{Model Formulation}
\label{sec:model}

\begin{table}[t]
    \centering
    \begin{tabular}{cl}
        \hline
        \textbf{Parameter} & \textbf{Definition}\\
        \hline
        $M$ & set of model configurations.\\
        $P$ & set of paths connecting IoT devices to cloud nodes.\\
        $C$ & set of compute nodes.\\
        $A$ & set of accuracy values achieved by model configurations.\\
        $f_{p,c}$ & equal to $1$ if path $p \in P$ terminates at compute node \\ & $c \in C$; $0$ otherwise.\\
        $l_{m,p}$ & latency introduced by path $p \in P$ when using model \\ & configuration $m \in M$.\\
        $l_{m,c}$ & latency introduced by inference with model configuration \\ & $m \in M$ executed at compute node $c \in C$.\\
        $\epsilon_{L}$ & maximum latency allowed.\\
        $e_{m, p}$ & energy consumption for path $p \in P$ for a single \\ & inference when using model configuration $m \in M$.\\
        $e_{m,c}$ & energy consumption of inference with model \\ & configuration $m \in M$ executed at compute node $c \in C$.\\
        $r_{m,c}$ & resources required to execute model $m \in M$ at \\ & compute node $c \in C$.\\
        $R_{c}$ & available resources at compute node $c \in C$.\\
        $a_{m}$ & average accuracy provided by model \\ & configuration $m \in M$.\\
        $\epsilon_{A}$ & required accuracy.\\
        \hline
        $x_{m,c}$ & binary variable equal to $1$ if configuration model \\ & $m \in M$ is selected to be executed at compute node \\ & $c \in C$; $0$ otherwise.\\
        $y_{m,p}$ & binary variable equal to $1$ if path $p \in P$ is \\ & selected to carry traffic for the considered UE using \\ & configuration model $m \in M$; $0$ otherwise.\\
        \hline
    \end{tabular}
    \vspace{2pt}
    \caption{Notation for the MO-ILP model and $\epsilon$-constrained method.}
    \label{tab:ILP-v2}
\end{table}

To account for energy consumption, latency, and accuracy in a joint fashion, the problem is formalized as \ac{MO}-\ac{ILP}. We first describe the considered objective functions. The first goal is to minimize the energy consumption of the end-to-end system:

\begin{equation}
	\label{obj_function_en}
	\text{min}\; f_{1} = \sum_{m\in{M}}\sum_{p\in{P}} y_{m,p}e_{m,p} + \sum_{m\in{M}}\sum_{c\in{C}} x_{m,c} e_{m,c}.
\end{equation}

The objective function \eqref{obj_function_en} is composed of two members. The first term takes into account the energy consumed by the network to transmit the data from an \ac{IoT} device to the selected compute node along the chosen path, while the second term accounts for performing an inference task on the selected compute node. This allows the joint selection of path and compute node that minimizes the end-to-end energy consumption. Each term is computed according to \ac{KPI} defined in Section \ref{sec:kpis}.
The second goal is to minimize the latency of the end-to-end system:

\begin{equation}
	\label{obj_function_lat}
	\text{min}\; f_{2} = \sum_{m\in{M}}\sum_{c\in{C}} x_{m,c}l_{m,c} + \sum_{m\in{M}}\sum_{p\in{P}} y_{m,p}l_{m,p}.
\end{equation}

The objective function \eqref{obj_function_lat} is analogous to the previous objective, but it considers the latency instead of energy.
The third goal is to maximize the accuracy of the end-to-end system:

\begin{equation}
	\label{obj_function_acc}
	\text{max}\; f_{3} = \sum_{m\in{M}}\sum_{c\in{C}} x_{m,c}a_{m}.
\end{equation}

The objective function \eqref{obj_function_acc} is the multiplication of the decision variable on the inference model and cloud location by the accuracy provided by the model. By considering that the accuracy of a model (see \eqref{eq:f1}) can be expressed as a value between $0$ and $1$, it is possible to define its complement $1-a_{m}$, representing the inaccuracy achieved by the inference model $m$. This transforms the maximization into a minimization problem with the following objective function:

\begin{equation}
	\label{obj_function_inacc}
	\text{min}\; \overline{f_{3}} = \sum_{m\in{M}}\sum_{c\in{C}} x_{m,c}(1-a_{m}).
\end{equation}

By jointly considering the three objective functions (\eqref{obj_function_en}, \eqref{obj_function_lat}, \eqref{obj_function_inacc}) the optimization problem is formulated as an \ac{MO}-\ac{ILP}:

\begin{equation}
	\label{obj_function}
	\text{min}\; (f_{1}, f_{2}, \overline{f_{3}})
\end{equation}

\begin{equation}
	\label{constraint1}
	\sum_{m\in{M}}\sum_{c\in{C}} x_{m,c} = 1
\end{equation}

\begin{equation}
	\label{constraint2-1}
	y_{m,p} \geq f_{p,c}x_{m,c}, \quad \forall m\in{M}, p\in{P}, c\in{C}
\end{equation}

\begin{equation}
	\label{constraint5}
	\sum_{m\in{M}} x_{m,c}r_{m,c} \leq R_{c}, \quad \forall c\in{C}.
\end{equation}


Constraint \eqref{constraint1} ensures that only one compute node is selected to perform the inference task.
Constraint \eqref{constraint2-1}, used in conjunction with the minimization in \eqref{obj_function_en}, ensures that only the path connecting the considered \ac{IoT} device to the chosen compute node is selected. This is done through parameter $f_{p,c}$, which simply contains the information on which paths can be used to connect the \ac{IoT} device to each cloud.
Constraint \eqref{constraint5} ensures that enough compute resources are available on the node selected to perform the inference task.

\subsection{Solving Multi-Objective Optimization Problems}

\ac{MO} optimization problems with conflicting objectives and no clear priorities are not straightforward to solve, as trade-offs must be found. For example, minimizing energy consumption might not minimize latency or accuracy. In general terms, a \ac{MO}-\ac{ILP} problem can be expressed as:
\begin{align}
\text{min} \quad & \mathbf{f}(\mathbf{x}) = \{f_{1}(\mathbf{x}),f_{2}(\mathbf{x}),...,f_{m}(\mathbf{x})\} \\
\text{s.t.} \quad & \mathbf{Ax} \leq \mathbf{b} \\
& \mathbf{x} \in \mathbb{Z}^n
\end{align}
where $(f_{1},f_{2},...,f_{m})$ are the objective functions, $\mathbf{x}$ is the decision variable vector, $\mathbf{A}$ is the constraint matix, and $\mathbf{b}$ the coefficients.
\ac{MO}-\ac{ILP} problems generally admit multiple optimal solutions. Solving such problems often requires determining the Pareto front, which captures the set of optimal trade-offs across the objectives \cite{Jones2022}. A feasible solution $\mathbf{x'} \in \mathbb{Z}^n$ is Pareto-optimal if there is no other feasible solution $\mathbf{x} \in \mathbb{Z}^n$ such that:

\begin{equation}
f_{i}(\mathbf{x}) \leq f_{i}(\mathbf{x'}) \; \forall i \in \{1,...,m\}
\end{equation}
and

\begin{equation}
f_{j}(\mathbf{x}) < f_{j}(\mathbf{x'}) \text{\, for at least one\, } j.
\end{equation}
%
%
The set of all Pareto-optimal solutions forms the Pareto set, and the corresponding set of objective vectors defines the Pareto front:

\begin{equation}
	PF = \{f(\mathbf{x}):\mathbf{x} \in \mathbb{Z}^n, \mathbf{x} \text{\, is Pareto-optimal}\}.
\end{equation}

Different approaches to enumerate or approximate the Pareto front can be used, with the $\epsilon$-constrained method being one of the most popular \cite{MAVAMC09}. In this method, one of the objectives is optimized while the others are converted into constraints bounded by $\epsilon$ values, which requires solving a sequence of single-objective \ac{ILP} problems. Formally, each problem to be solved can be expressed as follows:

\begin{align}
\text{min} \quad & f_{1}(\mathbf{x})\\
\text{s.t.} \quad & f_{2}(\mathbf{x}) \leq \epsilon_2\\
& \vdots \\
& f_{m}(\mathbf{x}) \leq \epsilon_m \\
& \mathbf{Ax} \leq \mathbf{b} \\
& \mathbf{x} \in \mathbb{Z}^n.
\end{align}

The $\epsilon$-constrained method can be applied to our problem by adding additional constraints. We keep the minimization of the energy consumption (i.e., $f_{1}$) as the main objective while we constrain latency and accuracy. For the latency, we provide a bound according to the \ac{KPI} in \eqref{eq:lat}:

\begin{equation}
	\label{constraint3}
	\sum_{m\in{M}}\sum_{c\in{C}} x_{m,c}l_{m,c} + \sum_{m\in{M}}\sum_{p\in{P}} y_{m,p}l_{m,p} \leq \epsilon_{L}.
\end{equation}

Constraint \eqref{constraint3} ensures that the sum of the inference latency (corresponding to $L_{\text{processing}}$) and latency along the selected path (corresponding to $L_{\text{radio}} + L_{\text{transport}} + L_{\text{routing}}$) does not exceed a maximum latency value $\epsilon_{L}$. For the accuracy, the bound is formulated as:

\begin{equation}
	\label{constraint4}
	\sum_{m\in{M}}\sum_{c\in{C}} x_{m,c}a_{m} \geq \epsilon_{A} \quad\forall \epsilon_{A} \in A.
\end{equation}

Constraint \eqref{constraint4} ensures that the accuracy of the selected model is at least equal to $\epsilon_{A}$, or larger. Based on this constraint, the algorithm selects the proper combination of inference tasks that satisfy the accuracy requirement. This constraint reflects the accuracy \ac{KPI} defined in \eqref{eq:f1}.


\section{Numerical Results}
\label{sec:results}

\subsection{Simulation Environment and Setup}

We developed a custom simulator written in Python to simulate a human detection system operating for $30$ days, assuming one image sent by each camera every $60$ seconds, for a total of $43200$ images. We generate images containing a human with a probability distributed according to a Poisson process. We refer to the mean value of such a distribution as \textit{event frequency}. All the results are an average of $1000$ simulation runs.

We consider the scenario depicted in Figure \ref{fig:sample-scenario}. For each considered strategy, i.e., \ac{TIoT}, \ac{IIoT-D}, and \ac{IIoT-C}, we assume a single \ac{IoT} device (also referred to as \ac{UE}), connected to a single \ac{AP} and, depending on the strategy, a single cloud node.
The consumption of a network interface card ($P_{\text{nic}}$) for backhaul traffic is considered to be equal to $1\, W$ \cite{power-model-future-proof}, while the consumption of a $100\, Gbps$ optical transponder ($P_{\text{tr}}$) is assumed to be $110.9\, W$ \cite{power-model-polimi}.
The \ac{IoT} device sends images either with resolution 640x640 with size ($d$) $1.25\, MB$ or 320x320  with size ($d$) $307\, kB$ and a transmitted power $P_t$ of $0.2\, W$ to the \ac{AP} every $60$ seconds.
Regarding the parameters for the \ac{AP} (needed to compute $P_{\text{AP}}^{\text{recv}}$ and $P_{\text{AP}}^{\text{proc}}$), we assume a $20\, MHz$ single-antenna, full load, spectral efficiency $6\, bps/Hz$, and $24$-bit quantization, with a maximum capacity of $1\, Gbps$. The power amplifier consumes $390\, W$ while the overhead factor $\eta$ is assumed to be $0.27$. We also assume that the device is served through a network slice providing it with a rate ($\alpha$) of $50\, Mbps$. The transport network is assumed to introduce a delay ($L_{\text{transport}}$) of $5\, ms$ for the edge cloud (two-hops away) and $168\, ms$ for the remote cloud ($15$-hops away). These values are within the range of real measurements on Amazon clouds available at \cite{cloud-ping} and already include the latency introduced by two \acp{UPF} ($N_{UPF}$).

\begin{table}[t]
    \centering
    \begin{tabular}{|c|c|c|c|}
        \hline
        \textbf{Category} & \textbf{Processor} & \textbf{GFLOPS} & \textbf{GHz} \\
        \hline
        Edge & Xeon Gold 6414U      & 1331 & 2.0 \\
        \hline
        Remote & Xeon Platinum  8480C  & 2329 & 2.0 \\
        \hline
        \multirow{2}{*}{Device}  
        & Raspberry Pi 5       & 31   & 2.4 \\
        & Intel Atom X6214RE   & 44   & 1.4 \\
        \hline
    \end{tabular}
    \vspace{2pt}
    \caption{CPU, GFLOPS, and frequency adopted in the simulations \cite{intel-cpu-gflops}\cite{intel-atom-gflops}\cite{other-gflops}.}
    \label{tab:cpus}
\end{table}
%
%

For the human detection task, we rely on \ac{YOLO}v10, a state-of-the-art deep learning model for real-time object detection \cite{yolo}\cite{yolo-code}. 
We consider different model sizes, depending on the number of parameters constituting the model, which translates into a different number of operations to perform the inference. Specifically, we consider $5$ sizes, namely N, S, M, L, X, which correspond to $6.61$, $21.71$, $68.5$, $87.6$, $195.9$ \ac{GFLOPS}.
We trained each model on the COCO dataset \cite{dataset-coco} for $50$ epochs with images with resolution 640x640 pixel, except for N, that we trained also to work with 320x320 image size, so to reduce the model operations down to $1.6$ \ac{GFLOPS}, thus making it more suitable to run on simple intelligent \ac{IoT} devices. In the following, we refer to the models with just 640 or 320 followed by a letter indicating the model size. For the \ac{TIoT} and \ac{IIoT-D} the threshold is set to $0.5$, so any inference score above it is considered as an image containing a human. For the \ac{IIoT-C} strategy, we set the cloud model threshold to $0.5$, as for the other strategies, and we vary the threshold for the model executed at the \ac{IoT} device. In the notation, we explicitly mention the latter threshold within square brackets and we omit the others, always set to $0.5$.
We consider different hardware configurations for cloud and \ac{IoT} computing devices, reported in Tab.\ref{tab:cpus}, for which we provide a performance analysis using the \acp{KPI} introduced in Section \ref{sec:use-case}. We consider different high-end Intel Xeon platforms for the cloud, while for the \ac{IoT} devices, which are usually equipped with simple hardware, we consider a Raspberry Pi 5 based platform and a more performing, efficient, and expensive platform based on an Intel Atom X6 designed for \ac{IoT}. For simplicity, we consider an effective switched capacitance of $k = 1.097 \cdot 10^{-27} [s/cycles]^3$ for all the processors \cite{BinTGCN23}.

\subsection{Simulation Results}

\begin{figure}[t]
    \centering
    \includegraphics[width=\linewidth]{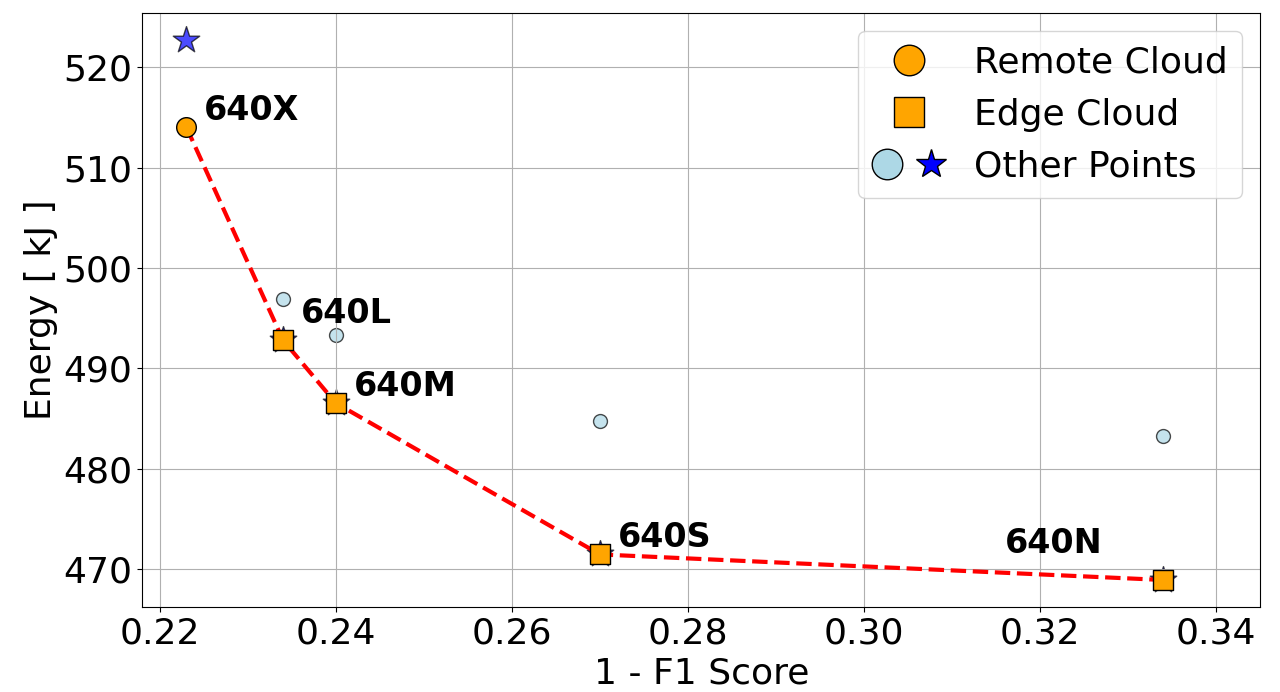}
    \caption{Total TIoT end-to-end energy consumption for different inference models as a function of inaccuracy. The solutions on the Pareto front for energy minimization are marked in orange.}
    \label{fig:res-TIoT-Energy}
\end{figure}

\begin{figure}[t]
    \centering
    \includegraphics[width=\linewidth]{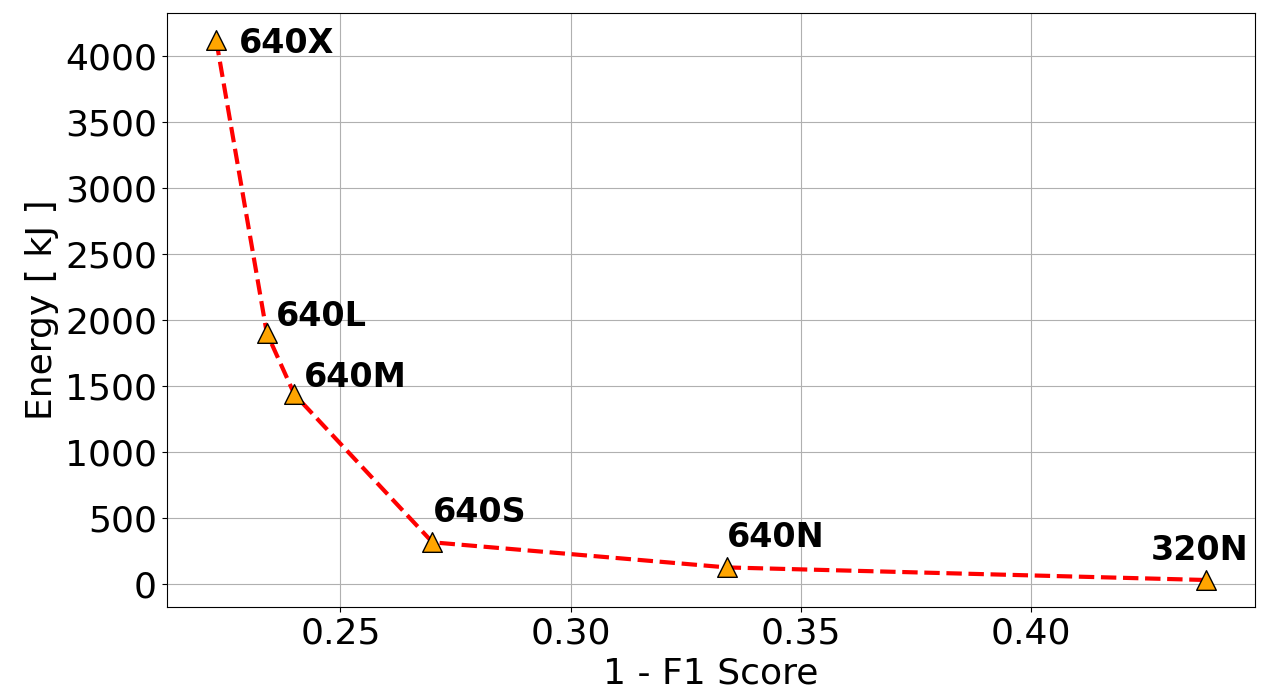}
    \caption{Total IIoT-D end-to-end energy consumption for different inference models as a function of inaccuracy. Devices are equipped with Raspberry Pi 5 hardware.}
    \label{fig:res-IIoT-D-Energy}
\end{figure}

\begin{figure}[t]
    \centering
    \includegraphics[width=\linewidth]{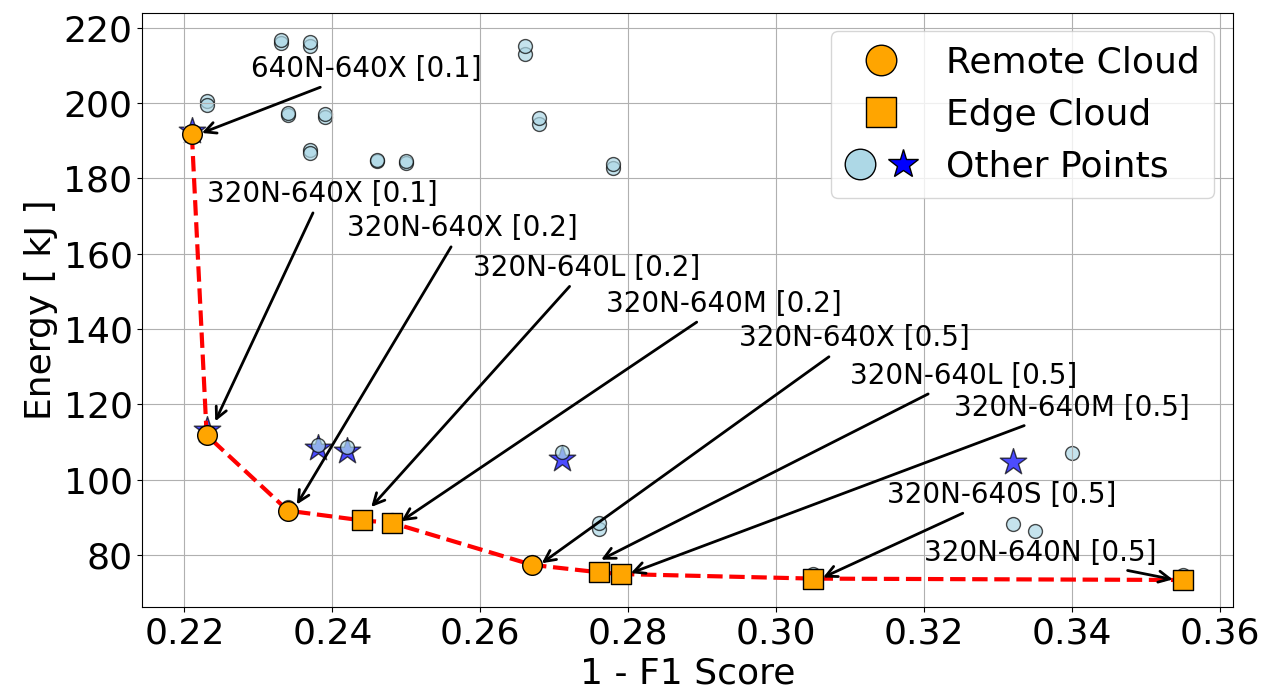}
    \caption{Total IIoT-C end-to-end energy consumption for different inference model combinations as a function of inaccuracy. The solutions on the Pareto front are marked in orange. Devices are equipped with Raspberry Pi 5 hardware.}
    \label{fig:res-IIoT-C-Energy}
\end{figure}

In the first set of results, we assume the event frequency to be $10\%$. 
As a first analysis, we consider the energy minimization (see \eqref{obj_function_en}) and show the results obtained with the Pareto front and epsilon-constrained method for the accuracy, while we relax the latency constraint. Figure \ref{fig:res-TIoT-Energy} depicts the total energy consumption as a function of the inaccuracy values during the whole simulation, using the \ac{TIoT} strategy and obtained when executing the inference with different models. It is possible to notice that the inaccuracy increases as the model size decreases, as models with a lower number of parameters provide less accurate predictions.
The energy consumption is also a function of the model, as models with a higher number of parameters and layers are more computationally expensive. From the figure, it is possible to notice that
the lowest cost solution for the largest model, 640X, is achieved when the inference is executed at the cloud. This is due to the fact that the 640X model requires a significant amount of computing power, and therefore it is more efficient to execute it on more powerful computing platforms in the remote cloud, even if some data must traverse long network paths. Conversely, for all other models, executing inference at the edge cloud is more energy efficient, as the energy required by the inference has less impact on the overall end-to-end energy consumption.
The figure also reports a series of points, which represent all the possible solutions (including dominated solutions). These points can be used to validate the optimization framework, as all the points that minimize the energy consumption for each inaccuracy value are correctly detected by the optimization model. Moreover, the points marked with a star represent the value of the energy consumption that one would obtain by applying the latency minimization strategy in \eqref{obj_function_lat} instead of the energy minimization. It is possible to notice that all solutions overlap with the Pareto front except for the 640X case, for which the solution on the remote cloud is not latency optimal. This suggests that in the \ac{TIoT} case, the energy consumption and latency minimization provide similar results and are not conflicting objectives. Conversely, targeting inaccuracy minimization or, equivalently, accuracy maximization, leads to large energy consumption, hence, the objectives are conflicting.

Figure \ref{fig:res-IIoT-D-Energy} is the equivalent plot to \ref{fig:res-TIoT-Energy} but obtained with the \ac{IIoT-D} strategy, where all the models are performed onboard the device with the Raspberry Pi 5 hardware configuration. The general trends are similar to the \ac{TIoT} case. However, while simple models like 320N and 640N can be easily executed onboard the devices, the energy consumption spikes for larger models. This is due to the fact that the Raspberry Pi 5 hardware platform does not execute inference tasks as efficiently as a cloud node equipped with a high-end \acp{CPU}.

Figure \ref{fig:res-IIoT-C-Energy} depicts the total energy consumption of the whole simulation as a function of the inaccuracy values and using the \ac{IIoT-C} strategy and Raspberry Pi 5 as \ac{IoT} device. For this analysis, we assume that \ac{IoT} devices can filter out irrelevant images only with low-complexity models (320N and 640N) suitable for low-performing hardware platforms. From the figure, it can be noticed that using low-complexity models and large thresholds results in low energy consumption but also high inaccuracy. When the threshold at the device side is set to $0.5$, many images are flagged as non-relevant and therefore not sent to the cloud for elaboration. Decreasing this threshold results in a larger number of images sent to the cloud, where more performing models can be executed, decreasing the inaccuracy at the expense of a higher energy consumption. 
When the high-complexity and high-performing 640X model is executed on the cloud, the optimization framework always selects to execute it on the remote cloud where the most efficient hardware platform is located, similarly to the \ac{TIoT} case. 
The combinations employing the 640X model in the cloud are also the most energy-consuming. 
Overall, the Pareto front allows for a qualitative analysis of the effects of service-level \acp{KPI}. Fixing a single specific requirement for inaccuracy might lead to energy-inefficient solutions. This is the case of the two left-most points on the curve, where a minimal inaccuracy difference is exhibited, but with one case requiring around $50\%$ more energy than the other. Conversely, the right side of the figure shows that with a small additional energy consumption, the system inaccuracy can be considerably improved. This suggests that adopting flexible service requirements can lead to a better overall optimization. This is the case of a service that might set a target \ac{KPI} with a range of acceptable degradation if the other \acp{KPI} can be improved by at least a minimum amount.
As a final note, the stars on the plot represent the solution to the optimization problem with a latency minimization objective instead of the energy consumption. For low and high inaccuracy values, the minimum latency solution overlaps with the minimum energy consumption. This is not the case for intermediate values, where minimizing the latency does not minimize the energy consumption and vice versa.

\begin{table*}[t]
\centering
\begin{tabular}{|l|c|c|c|c|c|}
\hline
\textbf{Component} & \textbf{TIoT 640X} & \textbf{IIoT-D RPi 640X} & \textbf{IIoT-D Atom 640X} & \textbf{IIoT-C RPi 640N-640X [0.1]} & \textbf{IIoT-C Atom 640N-640X [0.1]} \\
\hline
UE-proc & - & 4120.934 & 576.331 & 137.376 & 19.224 \\
UE-TX & 1.698 & - & - & 0.179 & 0.179 \\
AP (RAN) & 446.367 & - & - & 47.219 & 47.225 \\
TN & 17.976 & - & - & 1.901 & 1.901 \\
UPF & 16.243 & - & - & 1.718 & 1.718 \\
Cloud-proc & 31.752 & - & - & 3.358 & 3.359 \\
\hline
\textbf{Data sent} & 56.16 & 0 & 0 & 5.94 & 5.94 \\
\hline
\end{tabular}
\vspace{2pt}
\caption{Total energy consumption (in [kJ]) and data sent (in [GB]) for the different strategies and hardware platforms.}
\label{tab:energy_consumption}
\end{table*}

\begin{figure}[t]
    \centering
    \includegraphics[width=\linewidth]{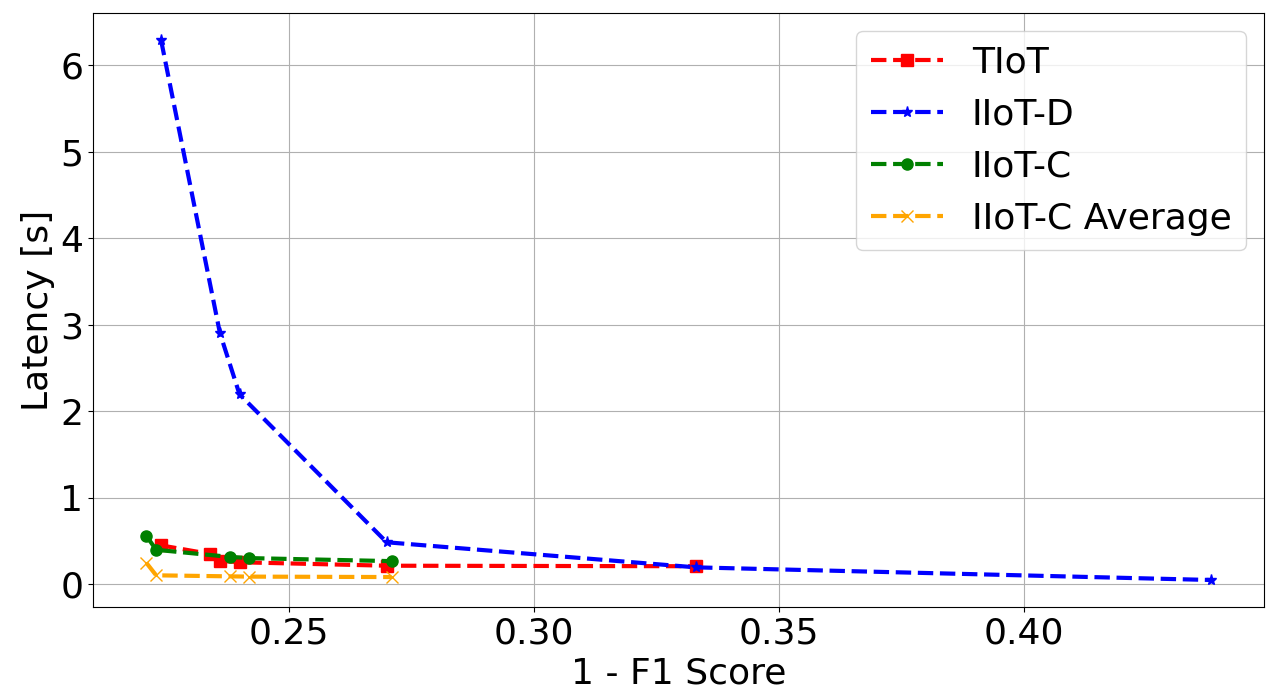}
    \caption{Pareto fronts for latency minimization for all strategies, as a function of inaccuracy. Devices are equipped with Raspberry Pi 5 hardware.}
    \label{fig:res-Latency}
\end{figure}

For an in-depth and fair comparison of the three strategies, let us now consider the case with the lowest inaccuracy, which corresponds to the first point on the left in each plot. Table \ref{tab:energy_consumption} shows the contributions to the end-to-end energy consumption for the different strategies with different \ac{IoT} hardware platforms in such conditions. In the \ac{TIoT} case, no elaboration is performed at the device and all images are sent to the cloud. In this case, more than $80\%$ of the energy is consumed by the \ac{AP} in the \ac{RAN} for wireless data transfer. A total of $56GB$ of data are sent over the network. In the \ac{IIoT-D} case, instead, all images are processed onboard the device, resulting in a very high energy consumption with the Raspberry Pi 5 hardware, while with the Atom platform the energy is more contained and similar to the \ac{TIoT} case. No images are sent over the network. The \ac{IIoT-C} case presents the lowest energy consumption. It is possible to appreciate how this approach is capable of reducing the energy consumption on the network, while slightly increasing the energy consumed to process images at the device. This is thanks to the models at the device that, despite their simplicity in terms of computational complexity, are able to filter out a great number of images directly at the source. Compared to the \ac{TIoT} strategy, the data sent over the network is reduced by $90\%$. Once again, the Atom platform outperforms the Raspberry Pi 5 in terms of energy efficiency.
It is worth noting that the Atom platform incurs a substantially higher cost compared to the Raspberry Pi. However, assessing this difference from a techno‑economic perspective is outside the scope of this paper.


Figure \ref{fig:res-Latency} shows the end-to-end latency needed to identify a human when using a Raspberry Pi 5 as \ac{IoT} device. The results are obtained with the Pareto front and epsilon-constrained method for the accuracy, while the energy is not constrained. These solutions correspond to the points marked with a star in the energy consumption plots. It is possible to see that the \ac{IIoT-D} case exhibits very high latency for low inaccuracy, due to the time needed to solve inferences with large models on a Raspberry Pi 5 platform. The latency of \ac{TIoT} and \ac{IIoT-D} is comparable and very close to each other, with a slight advantage for the former as no elaboration is performed at the \ac{IoT} device and the image is immediately sent to the cloud. In the \ac{IIoT-D} case, instead, the \ac{IoT} device always performs data elaboration. Finally, the figure also reports the \ac{IIoT-C} average latency. This case averages the latency presented above for \ac{IIoT-C} with the one where no humans are detected. This curve shows that, on average, the time needed to perform the inference is the lowest, as most of the time the device can identify images with no humans without the need for cloud intervention. For the \ac{TIoT} and \ac{IIoT-D} strategies, the curves also represent the average as they always employ a single model.

\begin{figure}[t]
    \centering
    \includegraphics[width=\linewidth]{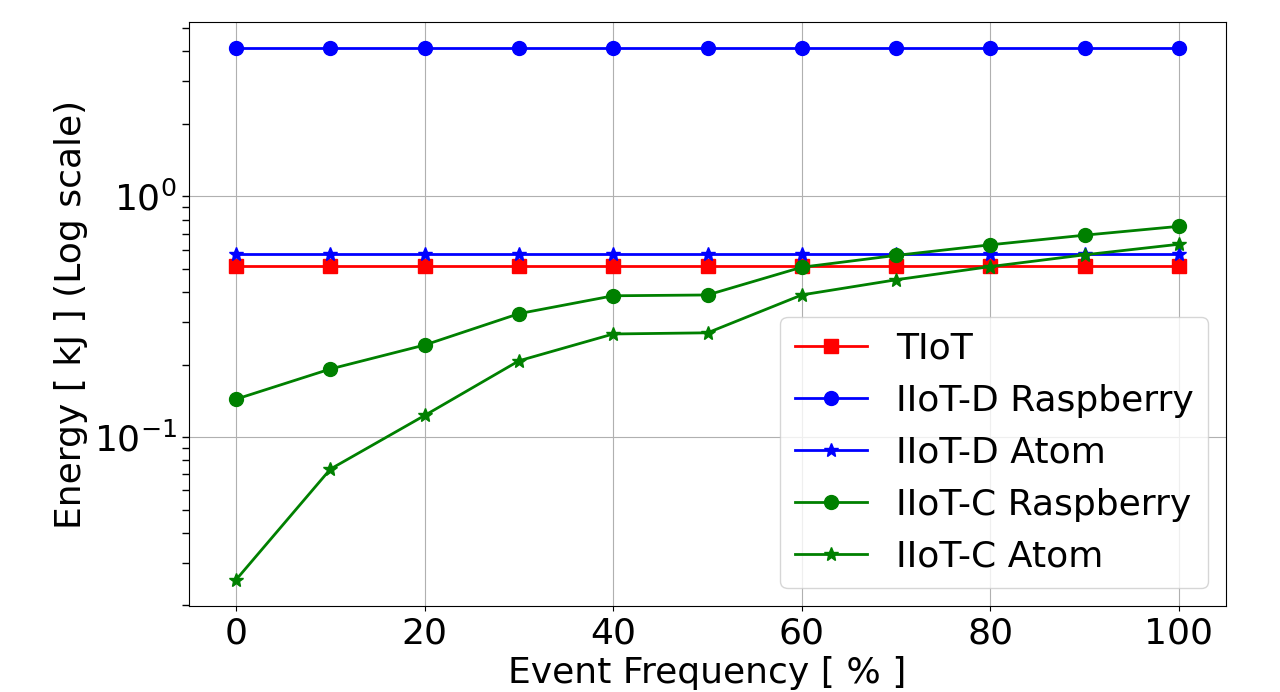}
    \caption{The effects of the event frequency on the total end-to-end energy consumption for all strategies.}
    \label{fig:all-Energy}
\end{figure}

So far, the results have been obtained for an event frequency fixed to $10\%$. Figure \ref{fig:all-Energy} depicts the overall end-to-end energy consumption during the simulation obtained with the different strategies and hardware platforms as a function of the event frequency. Similarly to Table \ref{tab:energy_consumption}, we focus on the case with the lowest inaccuracy value, which is obtained with the 640X model for \ac{TIoT} and \ac{IIoT-D}, and with 640N-640X with a threshold of $0.1$ at the device. From the plot, it can be noticed that the energy consumption for \ac{TIoT} and \ac{IIoT-D} does not depend on the event frequency, as these strategies always perform the same elaboration. The \ac{IIoT-C} instead depends on this frequency, as each image containing a human needs to be elaborated by the device and the cloud. The more images with humans, the higher the amount of data that needs to be sent and analyzed by the cloud, hence the higher the end-to-end energy consumption. The figure also shows that, with a Raspberry Pi 5 platform, the \ac{IIoT-C} strategy is convenient for event frequencies up to $60\%$, while with an Atom platform, benefits are experienced up to $80\%$. For higher values, almost all images contain humans, hence it is more efficient to perform single elaboration and use the \ac{TIoT} approach or the \ac{IIoT-D} if the hardware platform at the device is computationally efficient.

As a final analysis, we evaluated the computational complexity of the \ac{MO}-\ac{ILP} model. The model was solved using the \ac{ILP} solver Gurobi Optimizer version 12.0.0 \cite{gurobi} on a workstation equipped with an Intel i9-12900K CPU ($24$ cores) and $128$ $GB$ of RAM. Solving a single instance of the \ac{MO}-\ac{ILP} model requires less than $1s$. To assess scalability, we increased the problem size by considering up to $50$ models, cloud nodes, and paths, representing a large-scale scenario with randomly generated resource parameters. Across several runs, the solving time never exceeded $5s$.
This analysis demonstrates that the proposed model can be solved efficiently even for large-scale instances.

\section{Conclusion}
\label{sec:conclusion}

This paper focuses on assessing the end-to-end energy consumption, latency, and goal accuracy \acp{KPI} of \ac{GO} networking strategies. To this end, a practical framework is proposed that jointly considers network and service \acp{KPI}. An optimization model based on \ac{MO} \ac{ILP} is introduced to evaluate the trade-offs between the various \acp{KPI}. Simulation results show that the energy consumption and latency are heavily impacted by the service accuracy requirement to be provided with models of different complexity. Scenarios allowing for more inaccuracy are the ones where the most benefits can be achieved, as simple \ac{AI} models can be efficiently executed by \ac{IoT} devices or edge clouds, effectively offloading the network. On the contrary, when high accuracy is required, leveraging remote clouds is proven to be the most efficient solution. Results also suggest that, when frequent exchange of data is required, traditional \ac{IoT} solutions leveraging cloud resources are typically more efficient.
The methodology and framework developed in this paper enable a focused evaluation of the proposed \ac{GO} networking strategy within an object‑detection‑based intelligent \ac{IoT} scenario. While the study concentrates on this specific use case, it demonstrates how \ac{GO} principles can be applied in practice and lays the foundation for extending the approach to other tasks and modalities in future work.

\section*{Acknowledgment}
This work has been funded by the Huawei-CNIT Joint Innovation Center (JIC). The authors gratefully acknowledge Paolo Lanci for his contributions and collaboration during the early stages of this research.
\bibliographystyle{IEEEtran}
\bibliography{IEEEabrv,biblio}

\vfill

\end{document}